\documentclass{llncs}

\usepackage{amsfonts}
\usepackage{xcolor}
\usepackage{graphicx}
\usepackage{amssymb}
\usepackage{amsmath}
\usepackage{stmaryrd}

\usepackage{algorithm}
\usepackage{algpseudocodex}

\usepackage{hyperref}
\usepackage{cleveref}
\usepackage{xspace}
\usepackage{marginnote}
\usepackage{multirow}
\usepackage{rotating}
\usepackage{colortbl}
\usepackage{enumitem}
\usepackage{orcidlink} 
\usepackage[misc,geometry]{ifsym} 

\usepackage{caption}
\usepackage{subcaption} 

\usepackage{array}
\usepackage{booktabs}
\usepackage{soul} 
\usepackage{hhline}
\usepackage{enumitem}
\usepackage{stmaryrd}

\usepackage{mathtools}

\usepackage{adjustbox}
\usetikzlibrary{tikzmark, arrows.meta, positioning}

\usepackage{pgfplots}
\usepackage{tikz}

\usetikzlibrary{arrows,automata,fit,positioning,shadows,shapes}
\usetikzlibrary{decorations,snakes,calc} 
\usetikzlibrary{patterns, patterns.meta}
\tikzstyle{location}=[rectangle, rounded corners, minimum size=12pt, draw=black, fill=blue!10, inner sep=2pt]
\tikzstyle{pta}=[auto, ->, >=stealth']
\tikzstyle{PZG}=[auto, ->, >=stealth']
\tikzstyle{mergingFigure} = [>=stealth', node distance=1.8cm, yscale=.6]
\tikzstyle{location10}=[location, minimum size=10pt]
\tikzstyle{invariant}=[draw=black, dotted, inner sep=1pt, node distance=0] 
\tikzstyle{final}=[fill=green!70,double]
\tikzstyle{urgent}=[dotted, draw=red, very thick, fill=yellow]
\tikzstyle{bad}=[fill=red]

\tikzstyle{pzgstate} = [
	rectangle split,
	rectangle split horizontal,
	rectangle split parts=2,
	draw,
	rectangle split part align={center,left},
	rounded corners,
	minimum height=0.5 cm,
	minimum width=1.5 cm,
	thick
	]
\tikzstyle{fillred} = [ fill=red!20 ]
\tikzstyle{fillblue} = [ fill=blue!20 ]
\tikzstyle{fillyellow} = [ fill=yellow!20 ]
\tikzstyle{line} = [ draw,-latex',thick ]
\tikzstyle{highlightarrow} = [
	preaction={
		draw,
		line width=0.5mm
	}
]

\tikzstyle{urgent}=[fill=yellow, thick, dotted] 
\tikzstyle{private}=[fill=red!50,thick]

\pgfdeclarepatternformonly{north west lines wide}
{\pgfqpoint{-1pt}{-1pt}}
{\pgfqpoint{10pt}{10pt}}
{\pgfqpoint{9pt}{9pt}}
{
	\pgfsetlinewidth{3pt}
	\pgfpathmoveto{\pgfqpoint{0pt}{9.1pt}}
	\pgfpathlineto{\pgfqpoint{9.1pt}{0pt}}
	\pgfusepath{stroke}
}

\tikzset{onslide/.code args={<#1>#2}{%
		\only<#1>{\pgfkeysalso{#2}} 
}}

\definecolor{coloract}{rgb}{0.50, 0.70, 0.30}
\definecolor{colorclock}{rgb}{0.4, 0.4, 1}
\definecolor{colordisc}{rgb}{1, 0, 1}
\definecolor{colorloc}{rgb}{0.4, 0.4, 0.65}

\definecolor{colorparam}{rgb}{.66, 0.4, 0.0}
\definecolor{colorstate}{rgb}{1, 0.4, 0.0}

\definecolor{loccolor1}{rgb}{1, 0.6, 0.45}
\definecolor{loccolor2}{rgb}{0.45, 1, 0.45}
\definecolor{loccolor3}{rgb}{0.8, 0.8, 1}
\definecolor{loccolor4}{rgb}{1, 0.45, 1}
\definecolor{loccolor5}{rgb}{1, 1, 0.45}
\definecolor{loccolor6}{rgb}{0.45, 1, 1}
\definecolor{loccolor7}{rgb}{0.9, 0.6, 0.2}
\definecolor{loccolor8}{rgb}{0.7, 0.4, 1}
\definecolor{loccolor9}{rgb}{0.5, 1, 0.75}
\definecolor{loccolor10}{rgb}{0.8, 0.7, 0.6}
\definecolor{loccolor11}{rgb}{0.6, 0.7, 0.8}
\definecolor{loccolor12}{rgb}{0.2, 0.5, 0.9}
\definecolor{loccolor13}{rgb}{0.5, 0.9, 0.2}
\definecolor{loccolor14}{rgb}{0.9, 0.2, 0.5}
\definecolor{loccolor15}{rgb}{0.7, 0.7, 0.7}
\definecolor{loccolor16}{rgb}{0.8, 0.8, 0.5}

\usepackage{xcolor}
\definecolor{darkgreen}{rgb}{0.0, 0.4, 0.08}
\definecolor{lighterblack}{rgb}{.4, .4, .4}
\definecolor{colorparam}{rgb}{1, 0.6, 0.0}
\definecolor{mygreen}{rgb}{0,0.6,0}
\definecolor{mygray}{rgb}{0.5,0.5,0.5}
\definecolor{mymauve}{rgb}{0.58,0,0.82}

\definecolor{gris}{rgb}{0.6,0.6,0.6}
\definecolor{grisfonce}{rgb}{0.2,0.2,0.2}
\definecolor{turquoise}{rgb}{0, 1, 1}
\definecolor{vertfonce}{rgb}{0,0.85,0}
\definecolor{violet}{rgb}{0.8,0,0.8}
\definecolor{grispale}{rgb}{0.9, 0.9, 0.9}
\definecolor{cpale1}{rgb}{1, 0.3, 0.3}
\definecolor{cpale2}{rgb}{0.3, 1, 0.3}
\definecolor{cpale3}{rgb}{0.3, 0.3, 1}
\definecolor{cpale4}{rgb}{1, 0.3, 1}
\definecolor{cpale5}{rgb}{1, 1, 0.3}
\definecolor{cpale6}{rgb}{0.3, 1, 1}
\definecolor{cpale7}{rgb}{0.9, 0.6, 0.2}
\definecolor{cpale8}{rgb}{0.7, 0.4, 1}
\definecolor{cpale9}{rgb}{0.5, 1, 0.75}
\definecolor{cpale10}{rgb}{0.8, 0.7, 0.6}
\definecolor{cpale11}{rgb}{0.6, 0.7, 0.8}
\definecolor{cpale12}{rgb}{0.2, 0.5, 0.9}
\definecolor{cpale13}{rgb}{0.5, 0.9, 0.2}
\definecolor{cpale14}{rgb}{0.9, 0.2, 0.5}
\definecolor{cpale15}{rgb}{0.7, 0.7, 0.7}
\definecolor{cpale16}{rgb}{0.8, 0.8, 0.5}
\definecolor{bleuciel}{rgb}{0.90,0.95,1}
\definecolor{cv1}{rgb}{1, 0, 0}
\definecolor{cv2}{rgb}{0, 1, 0}
\definecolor{cv3}{rgb}{0, 0, 1}
\definecolor{cv4}{rgb}{1, 1, 0}
\definecolor{cv5}{rgb}{1, 0, 1}
\definecolor{cv6}{rgb}{0, 1, 1}
\definecolor{cv7}{rgb}{0.8, 0.6, 0.4}
\definecolor{cv8}{rgb}{0.5, 0.5, 1}
\definecolor{cv9}{rgb}{0.55, 0.75, 0.35}
\definecolor{cv10}{rgb}{1, 0.6, 0.1}
\definecolor{cv11}{rgb}{0.6, 0.7, 0.8}
\definecolor{cv12}{rgb}{0.2, 0.5, 0.9}
\definecolor{cv13}{rgb}{0.5, 0.9, 0.2}
\definecolor{cv14}{rgb}{1, 0.3, 0.5}
\definecolor{cv15}{rgb}{0.7, 0.7, 0.7}
\definecolor{cv16}{rgb}{0.8, 0.8, 0.5}
\definecolor{cvorange}{rgb}{1,.8,0.5}
\definecolor{colortask}{rgb}{0.5, 0.2, 0.9} 

\usepackage{xparse}

\newcommand{\marginX}{\marginnote{\huge{\quad\textbf{!}\quad}}}

\ifdefined\VersionWithComments
  \newcommand{\tocite}{\textcolor{red}{\marginX{}\sf[REF!]}\xspace}
  \newcommand{\todo}[1]{\textcolor{red}{\marginX{}\sffamily\bfseries TODO: #1}}
\else
  \newcommand{\tocite}[1]{}
  \newcommand{\todo}[1]{}
\fi

\newtheorem{requirement}{Requirement}

\crefname{definition}{Def.}{Defs.}
\crefname{theorem}{Thm.}{Thms.}
\crefname{lemma}{Lem.}{Lemmas}
\crefname{line}{Line}{Lines}
\crefname{algorithm}{Alg.}{Algs.}
\crefname{figure}{Fig.}{Figs.}
\crefname{appendix}{Appendix}{Appendices}
\crefname{section}{Sec.}{Sections}
\crefname{table}{Table}{Tables}
\crefname{requirement}{Req.}{Reqs.}

\newcommand{\ParamLinearTerm}{\ensuremath{\tau} }
\newcommand{\ZoneFormula}{\phi}
\newcommand{\comparator}{\ensuremath{\bowtie}}
\newcommand{\ComparatorSet}{\ensuremath{\mathtt{Cmp}}}
\newcommand{\subexpr}{\ensuremath{\mathtt{subexpr}}}

\newcommand{\ClockValSet}[1]{ \ensuremath{\mathbb{R}_{\geq 0}^{#1}}}
\newcommand{\ParamValSet}[1]{ \ensuremath{\mathbb{Q}_{\geq 0}^{#1}}}
\newcommand{\ValSet}{\ensuremath{V}}

\newcommand{\automaton}{\ensuremath{\mathcal{A}}}

\newcommand{\LocSet}{\ensuremath{L}}
\newcommand{\ClockSet}{\ensuremath{X}}
\newcommand{\ParamSet}{\ensuremath{P}}
\newcommand{\TransSet}{\ensuremath{T}}
\newcommand{\Inv}{\ensuremath{\mathit{Inv}}}

\newcommand{\GuardSet}{\ensuremath{\mathcal{G}}}

\NewDocumentCommand{\reset}{m o}{%
  \IfNoValueTF{#2}
    {Reset_{#1}}
    {Reset_{#1}(#2)}%
}

\newcommand{\genschema}{\ensuremath{\Sigma}}
\newcommand{\DistSet}{\ensuremath{\mathcal{D}}}
\newcommand{\Dist}{\ensuremath{\mu}}
\newcommand{\LocSetDist}{\ensuremath{\mathfrak{L}}}
\newcommand{\ParamSetDist}{\ensuremath{\mathfrak{P}}}
\newcommand{\ClockSetDist}{\ensuremath{\mathfrak{X}}}
\newcommand{\InvFunDist}{\ensuremath{\mathfrak{I}}}
\newcommand{\GuardSetDist}{\ensuremath{\mathfrak{Z}}}

\newcommand{\ClockResetDist}{\ensuremath{\mathfrak{Y}}}

\newcommand{\GraphDist}{\ensuremath{\mathfrak{G}}}

\newcommand{\MapDist}[1]{\ensuremath{#1_{\ast}}}

\newcommand{\modelspec}{\ensuremath{\mathcal{M}}}
\newcommand{\LocInterval}{\ensuremath{I_{\LocSet}}}
\newcommand{\ClockInterval}{\ensuremath{I_{\ClockSet}}}
\newcommand{\ParamInterval}{\ensuremath{I_{\ParamSet}}}
\newcommand{\OutdegInterval}{\ensuremath{I_{K}}}
\newcommand{\InvProb}{\ensuremath{p_{Inv}}}
\newcommand{\GuardProb}{\ensuremath{p_{g}}}
\newcommand{\ResetProb}{\ensuremath{p_{R}}}

\newcommand{\maxConst}{\ensuremath{N}}

\newcommand{\guard}{\ensuremath{g}}

\newcommand{\loc}{\ensuremath{\ell}}
\newcommand{\val}{\ensuremath{v} }

\newcommand{\imitator}{\textsc{Imitator}\xspace}

\newcommand{\PowerSet}[1]{\ensuremath{2^{#1}}}

\newcommand{\PosIntervalSet}{\ensuremath{\mathbb{I}_{\geq 1}}}
\newcommand{\IntervalSet}{\ensuremath{\mathbb{I}_{\geq 0}}}
\newcommand{\IntPts}[1]{\ensuremath{\mathbb{Z}(#1)}}
\newcommand{\interpret}[1]{\ensuremath{\llbracket #1 \rrbracket}}
\newcommand{\biginterpret}[1]{\ensuremath{\Bigl\llbracket #1 \Bigr\rrbracket}}
\newcommand{\Fin}{\ensuremath{\text{Fin}}}
\newcommand{\Uniform}{\ensuremath{\text{Uniform}}}
\newcommand{\Bern}{\ensuremath{\text{Bern}}}
\newcommand{\Choose}{\ensuremath{\text{Choose}}}
\newcommand{\Geom}{\ensuremath{\text{Geom}}}

\newcommand{\ChooseGeom}{\ensuremath{\text{ChooseGeom}}}

\newcommand{\bind}{\mathbin{\circledast}}

\newcommand{\Loc}{\ensuremath{\texttt{Loc}}}
\newcommand{\Clock}{\ensuremath{\texttt{Clock}}}
\newcommand{\Param}{\ensuremath{\texttt{Param}}}
\newcommand{\Tag}{\ensuremath{Tag}}

\crefformat{line}{#2l. #1#3}
\crefrangeformat{line}{ll. #3#1#4--#5#2#6}
\crefmultiformat{line}{ll. #2#1#3}{ and #2#1#3}{, #2#1#3}{ and #2#1#3}

\title{Random Generation of Small Quantitative Automata for Algorithm Debugging%
\thanks{This work was supported by Innovationsfonden Danmark's DIREC project SIoT (Secure Internet of Things).}%
\thanks{This preprint has not undergone peer review (when applicable) or any post-submission improvements or corrections. The Version of Record of this contribution is published in TASE 2026, and is available online at \url{https://doi.org/10.1007/978-3-032-30693-7_13}.}}

\author{
 Mikael Bisgaard Dahlsen-Jensen \Letter\orcidlink{0000-0003-0641-7635}\and
 Jaco van de Pol \orcidlink{0000-0003-4305-0625}
 }

\institute{
 Aarhus University, Aarhus, Denmark
 \\\email{mikael@cs.au.dk}
 }

 \authorrunning{Dahlsen-Jensen}

\titlerunning{}

\begin{document}
\maketitle

\setlength{\abovedisplayskip}{0pt}
\setlength{\belowdisplayskip}{0pt}
\begin{abstract}
    Analysis algorithms for quantitative automata are complex and hard to validate. Existing approaches --- benchmarks, mutation testing, uniform random generation --- each fail to expose subtle implementation bugs. We present a framework that repeatedly 1) generates random quantitative automata that are non-degenerate by construction, 2) tests each against a target property, and 3) shrinks any violation to a local minimum, yielding a small, actionable counterexample. We implement the framework for parametric timed automata (PTA) and apply it to \imitator, a mature model checker for PTA, uncovering 5 previously unknown bugs, one of which was exposed by a counterexample with just 2 locations and 1 transition.
\end{abstract}

\section{Introduction}
\label{sec:introduction}
Quantitative automata (timed, parametric, weighted, probabilistic) underpin a wide range of formal verification tools. Their analysis algorithms are complex and subject to continuous optimisation: as tools mature and acquire new features, subtle bugs can accumulate in corner cases that standard benchmark suites never exercise. Small, actionable counterexamples are crucial for debugging such bugs, but manually crafting them is difficult when the nature of the fault is unknown.

We propose a semantics-respecting random generation framework that avoids degenerate models---unsatisfiable guards, unreachable locations---by construction, paired with a shrinking procedure that iteratively removes structure while verifying the target property is preserved, yielding a small, human-readable witness. While we instantiate the framework for parametric timed automata (PTAs)~\cite{DBLP:conf/stoc/AlurHV93}, the design is not specific to them.

We demonstrate that the framework we propose produces them automatically: applying it to \imitator~\cite{DBLP:conf/cav/Andre21}, a mature model checker for PTAs, it uncovered 5 previously unknown bugs. One counterexample consists of just two locations and one transition, yet it exposes a fault in the unavoidability synthesis algorithm that had gone undetected until now.

Existing approaches fall short in different ways. Benchmark suites~\cite{DBLP:conf/tap/AndreMP21} of curated models rarely stress edge cases. Mutation testing requires a collection of good seeds, but any bias in those seeds risks missing entire classes of faults; and when the nature of the bugs is unknown, there is no principled basis for selecting them. Uniform random generation avoids this bias but produces mostly degenerate models with unsatisfiable constraints or unreachable locations that no algorithm can meaningfully distinguish.

\subsubsection{Related work}
Property-based testing~\cite{DBLP:conf/icfp/ClaessenH00} introduced the generate-and-shrink paradigm for automated counterexample finding.
Random generation to stress-test analysis tools is well-established: compiler testing~\cite{DBLP:conf/pldi/YangCER11} and SMT solver fuzzing~\cite{10.1145/1670412.1670413} demonstrate that random inputs reliably surface subtle implementation faults.
Random NFA generation~\cite{DBLP:conf/lpar/TabakovV05} has been used to benchmark algorithm performance rather than find bugs: performance evaluation tolerates degenerate inputs and has no need for small witnesses, goals orthogonal to ours.
Most closely related is the work of Manini et al.~\cite{DBLP:conf/tase/ManiniRP25}, who generate random timed automata and derive a problem-specific oracle automatically; however, constructing such an oracle requires non-trivial problem-specific theory, limiting applicability to problems where one is available, and they perform no counterexample minimisation.
We sidestep the oracle problem entirely via differential testing~\cite{DBLP:journals/dtj/McKeeman98}, comparing two algorithm variants on the same input, and reduce any witness to a minimal non-witness example through property-preserving shrinking.
 
\subsubsection{Contributions}
\begin{itemize}
\item A semantics-respecting random PTA generator that produces non-degenerate models by construction
\item A shrinker that reduces witnesses to a local minimum while preserving the target property, yielding small actionable counterexamples
\item An implementation as a meta-tool within \imitator
\item An empirical evaluation uncovering 5 previously unknown bugs in a mature tool

\end{itemize}

\section{Background}
\label{sec:background}

\subsection{Parametric Timed Automata}
A Parametric Timed Automaton (PTA) is a structure based on timed automata (TA). Like classical automata, it has locations connected by discrete transitions. It also has clocks.
Locations are associated a condition on clock valuations (invariant) that must
be satisfied while staying in the location.
An action in a timed automaton is either to take a discrete transition or to let some
time pass.
Discrete transitions have a guard that must be
satisfied in order to take the transition, and a set of clocks to reset. In a parametric setting, these conditions
use linear terms over clocks and parameters. Parameters are unspecified but constant during a run.

Throughout this paper, we let $\ClockSet$ denote a set of real-valued clocks and $\ParamSet$  a set of rational parameters.

\begin{definition}[parametric guards]
\label{def:parametric-guard}
A \emph{linear term} over $\ParamSet$ has the form $k$ or $kp$ or a sum thereof, where $k \in \mathbb{Q}$ and $p \in \ParamSet$.
The set of \emph{parametric guards} $\GuardSet(\ClockSet,\ParamSet)$ is defined by the grammar:
$$\ZoneFormula \; :=
\; \top
\; | \; \ZoneFormula \land \ZoneFormula \; | \; x \comparator \ParamLinearTerm
\; | \; \ParamLinearTerm' \comparator \ParamLinearTerm$$
where $x\in\ClockSet$, ${\comparator} \in \ComparatorSet := \{ <, \leq, =, \geq, > \}$, and $\ParamLinearTerm, \ParamLinearTerm'$ are linear terms over $\ParamSet$.
\end{definition}

In this paper, parametric guards are treated as syntactic objects. Their semantics will be defined in terms of the set of valuations that satisfy them (\cref{def:guard-satisfaction}).

\begin{definition}[atomic sub-expressions]
Let $\phi \in \GuardSet(\ClockSet,\ParamSet)$ be a parametric guard.
The multiset of \emph{atomic sub-expressions} of $\phi$, denoted $\subexpr(\phi)$,
is the multiset of sub-expressions of $\phi$ that are of the form
$x \comparator \ParamLinearTerm$ or
$\ParamLinearTerm' \comparator \ParamLinearTerm$.
\end{definition}

\begin{example}
Consider the parametric guard
$\phi = (x \leq p + 5) \land (y > 2p) \land (x \leq p + 5)$.
The multiset of atomic sub-expressions of $\phi$ is:
$$
\subexpr(\phi) = \{ x \leq p + 5, x \leq p + 5, y > 2p \}.
$$
\end{example}

\begin{definition}[Parametric Timed Automaton]
A \emph{Parametric Timed Automaton} is a tuple of the form
$\automaton = (\LocSet , \ClockSet, \ParamSet, \TransSet, \loc_0, \Inv)$ such that
\begin{itemize}[noitemsep, topsep=0pt]
    \item $\LocSet$, $\ClockSet$, $\ParamSet$  are disjoint sets of \emph{locations}, \emph{clocks},
        \emph{parameters}.
    \item
        $\TransSet \subseteq \LocSet \times \GuardSet(\ClockSet,\ParamSet) \times \PowerSet{\ClockSet} \times \LocSet$
        is the set of \emph{transitions} of the form
        $(\loc,\guard,Y,\loc')$ where:
        $\loc$, $\loc'$ are source
        and target locations, $\guard$ is the parametric guard, $Y$ the set of clocks to reset.
        We write $\loc \xrightarrow{\guard,Y} \loc'$ to denote a transition and $\loc \to \loc'$ when the guard and resets are unimportant.
    
    \item $\loc_0$ is the initial location.
    \item $\Inv \; : \; \LocSet \to \GuardSet(\ClockSet,\ParamSet)$ associates an
        \emph{invariant} with each location.
\end{itemize}
\end{definition}

\begin{definition}[valuations]
A \emph{clock valuation} is a function $\val_\ClockSet \in \ClockValSet{\ClockSet}$ assigning a non-negative real value to each clock in $\ClockSet$.
A \emph{parameter valuation} $\val_\ParamSet \in \ParamValSet
    {\ParamSet}$ assigns a non-negative rational value to each parameter in $\ParamSet$.

    A \emph{valuation} is a pair $\val = (\val_\ClockSet,\val_\ParamSet)$. 
    The set of all valuations of a PTA with clocks $\ClockSet$ and parameters $\ParamSet$ is denoted
    $\ValSet = \ClockValSet{\ClockSet} \times \ParamValSet{\ParamSet}$.
Since $\ClockSet$ and $\ParamSet$ are disjoint, we write $v(y)$ for 
$v_{\ClockSet}(y)$ if $y \in \ClockSet$ and $v_{\ParamSet}(y)$ if $y \in \ParamSet$.
\end{definition}

\begin{definition}[clock update]
\label{def:clock-update}
For $Y \subseteq \ClockSet$ and $\val \in \ValSet$, write $\val[Y \mapsto 0]$ for the valuation that maps each $x \in Y$ to $0$ and each $x \notin Y$ to $\val(x)$. For $Z \subseteq \ValSet$, write $Z[Y \mapsto 0]^{-1} := \{\val \mid \val[Y \mapsto 0] \in Z\}$, i.e., the set of valuations that land in $Z$ after resetting the clocks in $Y$.
\end{definition}

\begin{definition}[satisfaction of guards]
    \label{def:guard-satisfaction}
    Let $\guard \in \GuardSet(\ClockSet,\ParamSet)$ be a parametric guard, and $\val = (\val_\ClockSet,\val_\ParamSet)$ a valuation. We say that $\val$ \emph{satisfies} $\guard$, denoted $\val \models \guard$, if $\guard$ evaluates to true when each clock $x \in \ClockSet$ is replaced by $\val_\ClockSet(x)$ and each parameter $p \in \ParamSet$ is replaced by $\val_\ParamSet(p)$. Let $\interpret{\guard}$ denote the set of valuations satisfying $\guard$:
    $$\interpret{\guard} = \{\val \in \ValSet \mid \val \models \guard\}.$$
    
\end{definition}

\subsection{Multigraphs}
We recall the standard notion of directed multigraphs, which we use to describe the graph structure of generated automata.

\begin{definition}[multigraph]
A directed multigraph is a pair $G = (\LocSet, M)$, where
$L$ is a finite set of locations and
$M : \LocSet \times \LocSet \to \mathbb{N}$ is an edge multiplicity function.
For $\loc,\loc' \in L$, let the value $M(\loc,\loc')$ be the number
of directed edges from $\loc$ to $\loc'$.

The outgoing degree of a location $\loc \in \LocSet$ (counting multiplicities)
is defined as
$$
\deg^+_M(\loc) \;\triangleq\; \sum_{\loc' \in \LocSet} M(\loc,\loc').
$$    
\end{definition}

\subsection{Probability Distributions}

We use probability distributions to describe and parameterize the random generation of automata. 

\begin{definition}[Distribution]
For a countable set $A$, let $\DistSet(A)$ denote the set of probability distributions over $A$,
i.e.\ functions $\Dist : A \to [0,1]$ such that $\sum_{a \in A} \Dist(a) = 1$.
We write $X \sim \Dist$ to denote that $X$ is sampled according to $\Dist$.
\end{definition}

We now introduce several standard constructions on distributions that we use to describe our generation algorithms.

\begin{definition}[Standard distributions]
We use the following standard distributions on countable sets.
\begin{itemize}[noitemsep]
    \item $\Uniform(A)$: uniform over a non-empty finite set $A$, assigning probability $1/|A|$ to each element.
    \item $\Bern(p)$: Bernoulli, $p \in [0,1]$: taking $1$ with probability $p$ and $0$ with $1-p$.
    \item $\Choose_k(A)$: uniform over $k$-element subsets of finite set $A$, for $0 \leq k \leq |A|$.
    \item $\delta_a$ is the Dirac distribution: a point mass at $a$, with $\delta_a(a) = 1$.
    \item $\Geom_{\le m}(q)$: Capped Geometric, for $q \in (0,1)$ and $m \in \mathbb{N}_{\ge 1}$, the distribution on $\{1,\dots,m\}$ with $P(K{=}k) = (1-q)q^{k-1}$ for $k < m$ and $P(K{=}m) = q^{m-1}$.
\end{itemize}
\end{definition}

\begin{definition}[Pushforward]
Let $f : A \to B$ and $\Dist \in \DistSet(A)$.
The pushforward distribution $\MapDist{f}(\Dist) \in \DistSet(B)$ is defined by
$$
\MapDist{f}(\Dist)(b)
= \sum_{a \in A \,:\, f(a)=b} \Dist(a).
$$
Equivalently, if $X \sim \Dist$ then $f(X) \sim \MapDist{f}(\Dist)$.
\end{definition}

\begin{definition}[Dependent sampling]
Let $\Dist \in \DistSet(A)$ and $f : A \to \DistSet(B)$.
The distribution $\Dist \bind f \in \DistSet(B)$ is defined by
$$
(\Dist \bind f)(b)
=
\sum_{a \in A} \Dist(a)\, f(a)(b).
$$
Equivalently, sample $a \sim \Dist$ and then sample $b$ from $f(a)$.
\end{definition}

\begin{definition}[Product distribution]
Let $\Dist \in \DistSet(A)$ and $\nu \in \DistSet(B)$.
Their product $\Dist \otimes \nu \in \DistSet(A \times B)$ is defined by
$$
(\Dist \otimes \nu)(a,b) = \Dist(a)\nu(b).
$$
Equivalently, sample independently from $\Dist$ and $\nu$.
\end{definition}

\begin{definition}[$n$-fold product]
Let $\Dist \in \DistSet(A)$, $n \in \mathbb{N}$, and write $\Fin(n) := \{0,\dots,n-1\}$.
The $n$-fold product $\Dist^{\otimes n} \in \DistSet(A^{\Fin(n)})$ is defined by
$$
\Dist^{\otimes n}\bigl((a_i)_{i \in \Fin(n)}\bigr) := \prod_{i \in \Fin(n)} \Dist(a_i).
$$
Equivalently, $\Dist^{\otimes n}$ samples $n$ independent values from $\Dist$.
\end{definition}

\begin{definition}[Geometric subset choice]
For $q \in (0,1)$ and a finite set $S$, define
$$
\ChooseGeom(q,S) \;\triangleq\; \Geom_{\le |S|}(q) \;\bind\; (\lambda k.\; \Choose_k(S)).
$$
Equivalently, draw a random size $k$ from $\Geom_{\le |S|}(q)$ and return a uniformly random $k$-element subset of $S$.
\end{definition}

\section{Framework}
\label{sec:framework}
The framework consists of two components: a semantics-respecting random generator (\cref{sec:generator}) and a property-preserving shrinker (\cref{sec:shrinker}). Together they form the \textsc{Validator} (\cref{alg:validator}), searching for a small counterexample to a property $\varphi$.

\begin{algorithm}[b]
\caption{\textsc{Validator}: bounded random search for a minimal witness}
\label{alg:validator}
\begin{algorithmic}[1]
\Require Model specification $\modelspec$, property $\varphi$, bound $r \in \mathbb{N}$
\Ensure A minimal witness $\automaton$ with $\neg\varphi(\automaton)$, or $\bot$
\State $\genschema \gets \Call{ConstructSchema}{\modelspec}$
\For{$i \gets 1$ \textbf{to} $r$}
    \State $\automaton \sim \Call{AutomatonGen}{\genschema}$
    \If{$\neg\varphi(\automaton)$}
        \State \Return $\Call{Shrink}{\automaton, \varphi}$
    \EndIf
\EndFor
\State \Return $\bot$
\end{algorithmic}
\end{algorithm}

A \emph{model specification} $\modelspec$ is a user-supplied configuration controlling the intended size and structure of generated PTAs, and a \emph{generator schema} $\genschema$ is a structured collection of probability distributions over PTA components; both are defined formally in \cref{sec:generator}. The generator comprises two high-level procedures:
\textsc{ConstructSchema} (\cref{alg:constructschema}), which translates $\modelspec$ into $\genschema$, and
\textsc{AutomatonGen} (\cref{alg:automatongen}), which draws a PTA from $\genschema$.
The shrinker is \textsc{Shrink} (\cref{sec:shrinker}), which iteratively removes structure from a violating PTA while preserving the violation. \Cref{sec:generator} establishes formal requirements for the generator and describes the algorithms that satisfy them by construction; \cref{sec:shrinker} presents the staged shrinking pipeline and its design principles.

\section{Semantics-Respecting Generator}
\label{sec:generator}
This section develops the semantics-respecting PTA generator in three steps: we first state formal requirements that any meaningful generator must satisfy (\cref{sec:genreqs}), then introduce the generation scheme and \textsc{AutomatonGen}, which samples a PTA from a given schema (\cref{sec:genalgs}), and finally show how \textsc{ConstructSchema} builds a schema from a model specification so that all requirements are met by construction (\cref{sec:constructschema}).

\subsection{Generator Requirements}
\label{sec:genreqs}
We adopt a \emph{principle of meaningfulness}: all components of the generated PTA should be \textbf{meaningful}. This means that we want to avoid generating components that are trivial, unreachable, unsatisfiable, or redundant. We now outline the specific design goals we have for the generator.

\paragraph{Graph Structure}
The shape of the underlying graph of the generated PTA is an important aspect of generation.
A completely random graph structure has a high likelihood of being disconnected or having locations that are unreachable from the initial location. Following the principle of meaningfulness, this is unacceptable as it leads to meaningless locations. 

We therefore require all locations to be reachable from the initial location via discrete transitions.

\begin{requirement}[Graph Reachability]
\label{req:graphreach}
The underlying directed graph of the generated PTA is such that all locations are reachable from the initial location via discrete transitions, ignoring guards, invariants and time elapsing:
 $$\psi_{Graph}(\automaton = (\LocSet , \ClockSet, \ParamSet, \TransSet, \loc_0, \Inv)) \triangleq \forall \loc \in \LocSet, \loc_0 \to^* \loc $$ where $\to^*$ is the transitive closure of $\to$.
\end{requirement}

\paragraph{Invariants}
Invariants are important components of timed automata as they restrict the time that can be spent in a location. However, if an invariant is unsatisfiable, then the location becomes meaningless as no time can be spent there. Following the principle of meaningfulness, we want to avoid generating locations with unsatisfiable invariants.

\begin{requirement}[Invariant Satisfiability]
\label{req:invsat}
All invariants in the generated PTA are satisfiable:
 $$\psi_{InvSat}(\automaton = (\LocSet , \ClockSet, \ParamSet, \TransSet, \loc_0, \Inv)) \triangleq \forall \loc \in \LocSet, \interpret{\Inv(\loc)} \neq \emptyset$$
\end{requirement}

Furthermore, we want to ensure that the initial location invariant is satisfiable for the initial clock valuation where all clocks are set to 0. This ensures that the system can start executing from the initial location. In a sense, this is the most important requirement for invariants, as if this is not satisfied, the generated PTA is completely meaningless from the start.

\begin{requirement}[Initial Invariant Satisfiability]
\label{req:invinitsat}
The initial location invariant is satisfiable for the initial clock valuation where all clocks are set to 0:

 $$\psi_{InvInit}(\automaton = (\LocSet , \ClockSet, \ParamSet, \TransSet, \loc_0, \Inv)) \triangleq \biginterpret{\bigwedge\limits_{x\in X} x = 0} \subseteq \interpret{\Inv(\loc_0)}$$
\end{requirement}

\paragraph{Guards}
Similar to invariants, a guard is only meaningful if it is satisfiable, so this is a nice starting point. We can strengthen this even more: even if a guard is satisfiable for \emph{some valuations}, it might not be satisfiable in any \emph{context} where the guard is used i.e. on a transition between two locations with certain invariants. There are three components to consider here: the source location invariant, the guard itself, and the target location invariant (after resetting the appropriate clocks).
Thus, a first attempt at a meaningfulness requirement for guards is that each guard should be satisfiable within the context of the source location invariant and the target location invariant (after resetting the appropriate clocks):

\begin{align*}
    \psi_{GuardSatAttempt}(\automaton = (\LocSet , \ClockSet, \ParamSet, \TransSet, \loc_0, \Inv)) \triangleq
    \forall (\loc_1, \guard, Y, \loc_2) \in \TransSet,\; \\
\interpret{\guard} \cap \interpret{\Inv(\loc_1)} \cap \interpret{\Inv(\loc_2)}[Y \mapsto 0]^{-1} \neq \emptyset
\end{align*}

However, this requirement on guards imposes a significant restriction on invariants. Invariant generation is no longer a local problem, since now all pairs of invariants connected by a transition must have a non-empty intersection (taking into account resets). This makes invariant generation significantly more complex, as we can no longer generate invariants independently for each location. Instead, we must consider the entire graph structure and the transitions between locations when generating invariants, introducing a graph exploration aspect to invariant generation. A good middleground between meaningfulness and feasibility is to only consider the source location invariant when checking guard satisfiability. This way, we ensure that the guard is meaningful in the context of where it is used, without overly complicating invariant generation. Thus, we require the following:

\begin{requirement}[Guard Satisfiability]
\label{req:guardsat}
All guards in the generated PTA are satisfiable within the context of the source location invariant:
\begin{align*}
    \psi_{GuardSat}(\automaton = (\LocSet , \ClockSet, \ParamSet, \TransSet, \loc_0, \Inv)) \triangleq
    \forall (\loc_1, \guard, Y, \loc_2) \in \TransSet,\; \\
\interpret{\guard} \cap \interpret{\Inv(\loc_1)} \neq \emptyset
\end{align*}
\end{requirement}

\paragraph{Redundant sub-expressions}
A guard or invariant may contain redundant sub-expressions that do not contribute to the overall meaning of the constraint. For example, the guard $x \leq 5 \land x \leq 10$ contains the redundant sub-expression $x \leq 10$ since it is implied by $x \leq 5$. Following the principle of meaningfulness, we want to avoid generating guards and invariants with redundant sub-expressions. We define a helper predicate over $\GuardSet(\ClockSet, \ParamSet)$ to capture the absence of redundant sub-expressions:

$$
\psi_{NoRed}(\ZoneFormula) \triangleq \forall \ZoneFormula_1, \ZoneFormula_2 \in \subexpr(\ZoneFormula), \interpret{\ZoneFormula_1} \not\subseteq \interpret{\ZoneFormula_2}
$$

and then require that all guards and invariants in the generated PTA satisfy this property:

\begin{requirement}[No Redundant Sub-expressions]
\label{req:nored}
No guards or invariants in the generated PTA contain redundant sub-expressions:
\begin{align*}
    \psi_{Redundant}(\automaton = (\LocSet , \ClockSet, \ParamSet, \TransSet, \loc_0, \Inv)) \triangleq \; &\forall \loc \in \LocSet, \psi_{NoRed}(\Inv(\loc))
    \; \land \; \\ 
    &\forall (\loc_1, \guard, Y, \loc_2) \in \TransSet, \psi_{NoRed}(g)
\end{align*}
\end{requirement}

\subsection{Generation Scheme}
\label{sec:genalgs}
To describe the generator, we first introduce the notion of a generation schema, which is a collection of probability distributions over the components of a PTA. A generation schema defines how to sample each simple component of a PTA, such as the set of locations, clocks, parameters, the graph structure, invariants, guards, and resets. 
\begin{definition}[generation schema]
    A \emph{generation schema} is a tuple
$\genschema = (\LocSetDist,\ClockSetDist, \ParamSetDist, \GraphDist, \InvFunDist, \GuardSetDist, \ClockResetDist)$
where:
\begin{itemize}[noitemsep, topsep=0pt]
  \item $\LocSetDist$ is a probability distribution over finite sets of locations $\LocSet$,
  \item $\ClockSetDist$ is a probability distribution over finite sets of clocks $\ClockSet$,
  \item $\ParamSetDist$ is a probability distribution over finite sets of parameters $\ParamSet$,
  \item $\InvFunDist(\LocSet, \ClockSet,\ParamSet)$ is a probability distribution over invariant functions $\Inv : \LocSet \to \GuardSet(\ClockSet,\ParamSet)$,
  \item $\GraphDist(\LocSet)$ is a probability distribution over multigraphs over $\LocSet$,
  \item $\GuardSetDist(\ClockSet,\ParamSet)$ is a probability distribution over $\GuardSet(\ClockSet,\ParamSet)$,
  \item $\ClockResetDist(\ClockSet)$ is a probability distribution over $\PowerSet{\ClockSet}$.
\end{itemize}
\end{definition}

Since the types of locations, clocks, and parameters are not fixed, we use canonical tagged carriers whose sizes are supplied dynamically.
\begin{definition}[Tagged canonical carriers for generation]
\label{def:tagged-carriers}
Let $\Tag \coloneqq \{\Loc,\Clock,\Param\}$ be a set of tags. For each $\alpha\in\Tag$ and
$n \in \mathbb{N}$, define the tagged carrier $\Fin^{\Loc}(\alpha) := \Fin(n)\times \{\alpha\}$.
Distinct tags ensure pairwise disjointness, allowing unambiguous distinction of locations, clocks, and parameters in guards.
We write $\loc_n$, $x_n$, $p_n$ as shorthands for $(n,\Loc)$, $(n,\Clock)$, $(n,\Param)$; the initial location is fixed as $\loc_0 = (0,\Loc)$.
\end{definition}

\textsc{AutomatonGen} (\cref{alg:automatongen}) samples PTA components in dependency order.
The sets $\LocSet$, $\ClockSet$, $\ParamSet$ are drawn first, since all remaining distributions are parameterized by them.
Invariants are sampled before the transition structure, so that the already-sampled source invariant $\Inv(\loc)$ can be passed to $\GuardSetDist$ as context --- a dependency that an appropriately instantiated $\GuardSetDist$ can exploit to satisfy \cref{req:guardsat}.

\begin{algorithm}
\caption{\textsc{AutomatonGen}}
\label{alg:automatongen}
\begin{algorithmic}[1]
\Require Generation schema $\genschema = (\LocSetDist,\ClockSetDist, \ParamSetDist, \GraphDist, \InvFunDist, \GuardSetDist, \ClockResetDist)$
\Ensure A randomly generated automaton $\automaton$

\State $(\LocSet, \ClockSet, \ParamSet) \sim (\LocSetDist \otimes \ClockSetDist \otimes \ParamSetDist)$ \Comment{Sample components that parameterize the rest}

\State $\Inv \sim \InvFunDist(\LocSet, \ClockSet,\ParamSet)$ \Comment{$\InvFunDist$ is a distribution over functions $\LocSet \to \GuardSet(\ClockSet,\ParamSet)$}

\State $M \sim \GraphDist(\LocSet)$

\State $\TransSet \gets \emptyset$
\ForAll{$\loc,\loc' \in \LocSet$}
  \For{$i = 1$ to $M(\loc,\loc')$}
    \State $\guard \sim \GuardSetDist(\ClockSet,\ParamSet, \Inv, \loc)$ \Comment{$\GuardSetDist$ is a distribution over guards}
    \State $Y \sim \ClockResetDist(\ClockSet)$ \Comment{$\ClockResetDist$ is a distribution over subsets of $\ClockSet$}
    \State $\TransSet \gets \TransSet \cup \{(\loc, \guard, Y, \loc')\}$
  \EndFor
\EndFor
\Return $(\LocSet , \ClockSet, \ParamSet, \TransSet, \loc_0, \Inv)$
\end{algorithmic}
\end{algorithm}

The distributions in $\genschema$ are left abstract. The next subsection defines a model specification and the \textsc{ConstructSchema} procedure that instantiates them to satisfy the design goals by construction.

\subsection{Instantiation of Generation Schema}
\label{sec:constructschema}
We now instantiate the abstract generation schema: we define two sub-procedures, \textsc{GraphGen} and \textsc{GuardGen}, that enforce the structural requirements, and show how \textsc{ConstructSchema} uses them to assemble a complete, requirements-satisfying schema from a model specification.

\paragraph{Graph structure.}
To satisfy \cref{req:graphreach}, $\GraphDist$ must only produce graphs where all locations are reachable from $\loc_0$. \textsc{GraphGen} (\cref{alg:graphgen}) achieves this by first building a spanning tree rooted at $\loc_0$ (guaranteeing reachability), then adding random edges until the desired per-location outdegree is met.

\begin{algorithm}
\caption{\textsc{GraphGen}}
\label{alg:graphgen}
\begin{algorithmic}[1]
\Require Set of locations $\LocSet$ and an outdegree mapping $K : \LocSet \to \mathbb{N}_{\geq 1}$
\Ensure A directed graph multigraph $G=(\LocSet,M)$ where all locations are reachable from $\loc_0$
\State $A \gets \{\loc_0\}$ \Comment{locations already in the spanning tree}
\State $M := \{(u,v) \mapsto 0 \; | \; u,v \in \LocSet\}$ \Comment{$M : \LocSet \times \LocSet \to \mathbb{N}$}
\ForAll{$\loc \in \LocSet \setminus \{\loc_0\}$} \Comment{build spanning tree from $\loc_0$}
  \State $\loc' \sim \Uniform(A)$;\; $M(\loc',\loc) \gets 1$;\; $A \gets A \cup \{\loc\}$
\EndFor
\ForAll{$\loc \in \LocSet$}
  \While{$\deg^+_M(\loc) < K(\loc)$}
    \State $\loc' \sim \text{Uniform}(\LocSet)$
    \State $M(\loc,\loc') \gets M(\loc,\loc') + 1$
  \EndWhile
\EndFor
\State $G = (\LocSet,M)$
\State \Return $G$

\end{algorithmic}
\end{algorithm}

\paragraph{Guard generation.}
Guards must intersect the invariant of their source location (\cref{req:guardsat}), but a uniformly random guard is almost certainly disjoint from any fixed invariant. We therefore restrict generation to \emph{Coupled Interval Guards} (CIGs), a class of guards structured as interval-and-coupling triples that admit efficient witness-based sampling guaranteeing intersection by construction.

Throughout this section, an \emph{interval} denotes a closed, open, or half-open interval with non-negative integer endpoints: $[l,u]$, $(l,u]$, $[l,u)$, or $(l,u)$ with $l,u\in\mathbb{Z}_{\geq 0}\cup\{+\infty\}$.
Write $\IntPts{I}$ for the set of non-negative integers in $I$; for example, $\IntPts{[2,5)} = \{2,3,4\}$ and $\IntPts{(3,4)} = \emptyset$.

\begin{definition}[Coupled Interval Guard (CIG)]
Let $\ClockSet$ be a finite set of clocks and $\ParamSet$ a finite set of parameters.
A \emph{Coupled Interval Guard} (CIG) over $(\ClockSet,\ParamSet)$ is a triple
\[
C = (B_X, B_P, \kappa)
\]
where:
\begin{itemize}
    \item $B_X : \ClockSet \to \IntervalSet$ assigns to each clock $x \in \ClockSet$
    an interval $B_X(x)$ (closed, open, or half-open)
    with $l_x,u_x \in \mathbb{Z}_{\geq 0} \cup \{+\infty\}$,

    \item $B_P : \ParamSet \to \IntervalSet$ assigns to each parameter $p \in \ParamSet$
    an interval $B_P(p)$ (closed, open, or half-open)
    with $l_p,u_p \in \mathbb{Z}_{\geq 0} \cup \{+\infty\}$,

    \item $\kappa : \ClockSet \rightharpoonup (\ParamSet \times \ComparatorSet)$
    is a partial function (the \emph{coupling map}),
    where $\ComparatorSet = \{<,\le,>,\ge\}$.
\end{itemize}

\end{definition}

\paragraph{Interpretation.}
Every CIG $C = (B_X,B_P,\kappa)$ induces a parametric guard
$\ZoneFormula_C \in \GuardSet(\ClockSet,\ParamSet)$
by interpreting its components as conjunctions of atomic constraints.
Hence CIGs form a syntactic subclass of parametric guards. 

In particular, CIGs restrict the guard language to:
(i) interval bounds over individual clocks and parameters (closed endpoints give $\le$/$\ge$ constraints; open endpoints give $<$/$>$ constraints), and
(ii) at most one clock-parameter comparison per clock.
No clock-clock comparisons and no comparisons between linear parameter
terms are permitted.
From now on, we freely identify a CIG with its induced parametric guard.

\begin{example}
Let $\ClockSet = \{x\}$, $\ParamSet = \{p\}$, and let $C = (B_X, B_P, \kappa)$ with
$B_X(x) = [0,5)$, $B_P(p) = [0,3]$, and $\kappa(x) = (p,<)$.
The induced guard is $x \ge 0 \land x < 5 \land p \ge 0 \land p \le 3 \land x < p$.
We use this CIG as a running example below.
\end{example}

A CIG $C$ is \emph{integer-satisfiable} if $\llbracket C \rrbracket$ contains a witness whose clock and parameter values are all non-negative integers.
Note that for open intervals with integer endpoints, integer-satisfiability is strictly stronger than real-satisfiability: $(3 < x < 4)$ is satisfiable over $\mathbb{R}$ but not integer-satisfiable.
Throughout, we require all CIGs arising in the generation pipeline to be integer-satisfiable.

\begin{definition}[tight CIG]
A CIG $C = (B_X,B_P,\kappa)$ is \emph{tight} if
\begin{itemize}
    \item for every clock $x \in \ClockSet$ : $\IntPts{B_X(x)} = \{\, v(x) \mid v \text{ is an integer witness of } C \,\},$
    \item for every parameter $p \in \ParamSet$ : $\IntPts{B_P(p)} = \{\, v(p) \mid v \text{ is an integer witness of } C \,\}.$
    
\end{itemize}
Intuitively, the integer points of each box exactly span the values the variable takes across all integer witnesses of $C$.
\end{definition}

\begin{example}
The CIG $C$ from the previous example is not tight.
The strict coupling $x < p$ and the bound $p \le 3$ force $v(x) \le 2$ for every integer witness (since $v(x) < v(p) \le 3$ and $v(x) \in \mathbb{Z}_{\geq 0}$),
so $\{v(x) \mid v \text{ is an integer witness of } C\} = \{0,1,2\} \ne \{0,1,2,3,4\} = \IntPts{B_X(x)}$.
\end{example}

\paragraph{Tightening.}
Given a CIG $C=(B_X,B_P,\kappa)$, we write $\mathrm{tighten}(C)$ for a \emph{bound-tightened} form of $C$ whose integer witnesses are exactly those of $C$.
It is defined whenever $C$ is integer-satisfiable, and the result is itself integer-satisfiable: any integer witness of $C$ satisfies the coupling constraints and therefore also satisfies the tightened bounds.
In our restricted setting, tightening is local and bidirectional: for each coupled clock with $\kappa(x)=(p,\comparator)$, the bound of $B_X(x)$ in the direction of $\comparator$ is tightened to match the corresponding bound of $B_P(p)$ (closed endpoint for $\le$/$\ge$, open for $<$/$>$); symmetrically, the opposing bound of $B_P(p)$ is tightened to exclude parameter values for which no compatible integer clock value exists.

\begin{example}
Tightening $C$ yields $\mathrm{tighten}(C) = (B'_X, B'_P, \kappa)$ with $B'_X(x) = [0,3)$ and $B'_P(p) = [1,3]$.
In the forward direction, $\kappa(x)=(p,{<})$ propagates $p$'s upper bound $3$ as an open upper endpoint on $x$.
In the backward direction, $p = 0$ is excluded from $B'_P$ because no integer $x \ge 0$ satisfies $x < 0$; the smallest parameter value with a compatible integer clock is $p = 1$ (witnessed by $x = 0$).
\end{example}

\paragraph{Witness generation.}
Let $C=(B_X,B_P,\kappa)$ be integer-satisfiable.
$\textsc{WitnessGen}(C)$ denotes the distribution over integer witnesses $v=(v_X,v_P)\in\llbracket C\rrbracket$ defined by first computing $C'=\mathrm{tighten}(C)$, then drawing each $v_P(p)$ uniformly from $\IntPts{B'_P(p)}$, and each $v_X(x)$ uniformly from the integers in the (possibly coupling-restricted) interval $B'_X(x)$.
The joint tight condition on $B'_P$ ensures that every sampled $n_p \in \IntPts{B'_P(p)}$ has at least one compatible integer $n_x$ in the coupling-restricted interval, so the distribution is well-defined;
$v \in \llbracket C \rrbracket$ holds by construction, and every integer witness has positive probability.

\textsc{GuardGen} (\cref{alg:guardgen}) generates guards satisfying \cref{req:invsat,req:invinitsat,req:guardsat,req:nored} by construction.
It samples a witness valuation $v = (v_X,v_P) \in \llbracket C \rrbracket$ via $\textsc{WitnessGen}(C)$ (which tightens $C$ internally).
Using this witness, a \emph{maximal candidate} $C_{\max}$ is constructed:
for each clock and parameter, an interval (with randomly chosen open or closed endpoints) is sampled around the corresponding witness value, and for each clock a clock-parameter comparison is chosen such that it is satisfied by $v$.
Since $v$ is an integer witness, any combination of open or closed endpoints is valid: an open lower bound samples strictly below $v$, and an open upper bound samples strictly above $v$, so $v$ always lies in the resulting interval.
By construction, every atomic constraint in $C_{\max}$ is satisfied by $v$.

The final guard is obtained by sampling a geometrically sized random
subset of $\subexpr(C_{\max})$.
Since all atomic constraints in $C_{\max}$ are satisfied by $v$,
any conjunction of a subset of them is also satisfied by $v$.
Hence, for the returned guard $G$, $\llbracket C \rrbracket \cap \llbracket G \rrbracket \neq \emptyset.$
In particular, when the procedure is instantiated with the invariant
of the source location as seed CIG, \cref{req:guardsat} holds.

When instantiated with the trivial seed guard $\top$,
the procedure produces satisfiable guards,
establishing \cref{req:invsat}.
Instantiating the initial location invariant with $\bigwedge_{x \in \ClockSet} x = 0$ ensures that the generated invariant intersects the all-zeros valuation, establishing \cref{req:invinitsat}.
Finally, the guards induced by CIGs contain no redundant sub-expressions, since they are conjunctions of atomic constraints that are not implied by each other, ensuring \cref{req:nored}.

\begin{algorithm}
\caption{\textsc{GuardGen}}
\label{alg:guardgen}
\begin{algorithmic}[1]
\Require $\ClockSet$, $\ParamSet$, $\maxConst \ge 1$, integer-satisfiable CIG $C=(B_X,B_P,\kappa)$
\Ensure Parametric guard $G$ with $\llbracket C \rrbracket \cap \llbracket G \rrbracket \neq \emptyset$
\State $v \sim \textsc{WitnessGen}(C)$
\ForAll{$y \in \ClockSet \uplus \ParamSet$} \Comment{build $B^{\max}$; $1$ = open endpoint, $0$ = closed}
\State $t_l, t_u \sim \Bern(1/2) \otimes \Bern(1/2)$
\State $l_y \sim \Uniform(\{0,\ldots,v(y) - t_l\})$ \Comment{if $v(y)=0$, force $t_l \gets 0$}
\State $u_y \sim \Uniform(\{v(y) + t_u,\ldots,\maxConst\})$ \Comment{if $v(y)=\maxConst$, force $t_u \gets 0$}
\State $B^{\max}(y) \gets$ interval $[l_y, u_y]$ with endpoint types $t_l, t_u$
\EndFor
\ForAll{$x \in \ClockSet$} \Comment{build $\kappa^{\max}$}
\State $p \sim \Uniform(\ParamSet)$
\State $\comparator \sim \Uniform(\{\comparator \in \ComparatorSet \mid v(x) \comparator v(p)\})$
\State $\kappa^{\max}(x) \gets (p, \comparator)$
\EndFor
\State $C_{\max} \gets (B^{\max}|_{\ClockSet},\; B^{\max}|_{\ParamSet},\; \kappa^{\max})$
\State \Return $\bigwedge\limits_{\phi \in \ChooseGeom(0.5,\subexpr(C_{\max}))} \phi$
\end{algorithmic}
\end{algorithm}

\paragraph{Model specification.}
A \emph{model specification} is a user-supplied configuration controlling the intended size and density of generated PTAs; it is the input to \textsc{ConstructSchema}. Here $\PosIntervalSet$ denotes the set of non-empty closed intervals $[a,b]$ with $a,b\in\mathbb{N}_{\geq 1}$, used to specify size ranges.

\begin{definition}[model specification]
    A \emph{model specification} is a tuple $\modelspec = (\LocInterval, \ClockInterval, \ParamInterval, \OutdegInterval, \InvProb, \GuardProb, \ResetProb, \maxConst)$ where:
    \begin{itemize}[noitemsep, topsep=0pt]
        \item $\LocInterval \in \PosIntervalSet$ is the size interval for locations,
        \item $\ClockInterval \in \PosIntervalSet$ is the size interval for clocks,
        \item $\ParamInterval \in \PosIntervalSet$ is the size interval for parameters,
        \item $\OutdegInterval \in \PosIntervalSet$ is the size interval for outgoing transitions per location,
        \item $\InvProb \in [0,1]_{\mathbb{R}}$ is the probability of a non-trivial location invariant,
        \item $\GuardProb \in [0,1]_{\mathbb{R}}$ is the probability of a non-trivial transition guard,
        \item $\ResetProb \in [0,1]_{\mathbb{R}}$ is the probability of a clock reset on a transition,
        \item $\maxConst \in \mathbb{N}_{\geq 1}$ is the maximum constant in linear terms.
    \end{itemize}
\end{definition}

\paragraph{Schema assembly.}
\textsc{ConstructSchema} (\cref{alg:constructschema}) assembles the component distributions into a generator schema $\genschema$, wiring each parameter of $\modelspec$ to the appropriate distribution constructor.

\begin{algorithm}
\caption{\textsc{ConstructSchema}}
\label{alg:constructschema}
\begin{algorithmic}[1]
\Require Model specification $\modelspec = (\LocInterval, \ClockInterval, \ParamInterval, \OutdegInterval, \InvProb, \GuardProb, \ResetProb, \maxConst)$
\Ensure A generator schema $\genschema$

\State $\LocSetDist, \ClockSetDist, \ParamSetDist \gets \MapDist{\Fin^{\Loc}}(\Uniform(\LocInterval)), \MapDist{\Fin^{\Clock}}(\Uniform(\ClockInterval)), \MapDist{\Fin^{\Param}}(\Uniform(\ParamInterval))$
\State $\GraphDist \gets \lambda \LocSet.\;
    \Uniform(\OutdegInterval)^{\otimes |\LocSet|}
    \bind (\lambda K.\ \textsc{GraphGen}(\LocSet,K))$
\State $I \gets \lambda X,P.\; \Bern(\InvProb) \bind [0 \mapsto \delta_{\top}, 1 \mapsto \textsc{GuardGen}(\ClockSet,\ParamSet, \maxConst, \top)]$
\State $I_{0} \gets \lambda \ClockSet,\ParamSet.\; \Bern(\InvProb) \bind [0 \mapsto \delta_{\top}, 1 \mapsto \textsc{GuardGen}(\ClockSet,\ParamSet, \maxConst, \bigwedge_{x \in \ClockSet} x = 0)]$
\State $\InvFunDist \gets \lambda \LocSet, \ClockSet,\ParamSet.\; I_{0}(\ClockSet,\ParamSet) \mathop{\otimes} I(\ClockSet,\ParamSet)^{\otimes(|\LocSet|-1)}$
\State $\GuardSetDist \gets \lambda \ClockSet,\ParamSet, \Inv, \loc. \; \Bern(\GuardProb) \bind [0\mapsto\delta_{\top}, 1\mapsto\textsc{GuardGen}(\ClockSet,\ParamSet, \maxConst, \Inv(\loc))]$
\State $\ClockResetDist \gets \lambda \ClockSet.\;
    \Bern(\ResetProb) \bind [0\mapsto \delta_{\emptyset}, 1 \mapsto \ChooseGeom(1/2, \ClockSet)]$
\State \Return $(\LocSetDist,\ClockSetDist, \ParamSetDist, \GraphDist, \InvFunDist, \GuardSetDist, \ClockResetDist)$
\end{algorithmic}
\end{algorithm}

The set distributions $\LocSetDist, \ClockSetDist, \ParamSetDist$ sample finite sets from the specified size intervals. The graph distribution $\GraphDist$ first samples per-location outdegrees, then delegates to \textsc{GraphGen}. The invariant distributions encode the initial-location constraint (\cref{req:invinitsat}) by instantiating \textsc{GuardGen} with $\bigwedge_{x \in \ClockSet} x = 0$ at the initial location and $\top$ elsewhere. The guard distribution instantiates \textsc{GuardGen} with the invariant of the source location $\Inv(\loc)$, ensuring guard--invariant intersection (\cref{req:guardsat}) by construction. Clock resets complete the schema.

\section{Property-Preserving Shrinking}
\label{sec:shrinker}
Once a PTA violating the target property has been identified, the framework applies a shrinking phase inspired by delta debugging~\cite{DBLP:conf/esec/Zeller99} and test-case reduction~\cite{DBLP:conf/pldi/RegehrCCEEY12}: the model is iteratively reduced while verifying that the target property is preserved. The goal is to quickly obtain a small and comprehensible witness suitable for inspection and debugging, rather than to compute a formally minimal automaton.

The shrinker is intentionally agnostic to the nature of the target property, which is treated as an external oracle. Unlike generation, which enforces global semantic constraints by construction, shrinking is inherently heuristic and oracle-guided. We therefore present the shrinker in terms of design principles and a staged shrinking procedure.

\subsection{Shrinker Design Goals}
We adopt a \emph{principle of parsimony}: the shrinker aims to eliminate components of a witness PTA that are not relevant to satisfying the target property. Parsimony is not understood as formal minimality, but as a practical criterion for removing unnecessary structure while preserving the property of interest.

\paragraph{Preservation of Target Property}
Every accepted shrinking step must preserve the target property; a shrunk PTA that no longer witnesses the violation is useless.

\paragraph{Monotonic Progress}
Each shrinking step strictly decreases a chosen measure of model size or complexity, such as the number of locations, transitions, clocks, or constraints. This guarantees progress and termination, and enforces parsimony by ensuring each accepted step removes structure rather than rearranging it.

\paragraph{Well-Formedness Preservation}
All shrinking operators preserve well-formedness of parametric timed automata. In particular, the shrinker avoids introducing ill-defined models, such as PTAs with inconsistent clock constraints or syntactically invalid components. This ensures parsimony does not produce degenerate or semantically meaningless witnesses.

\paragraph{Locality and Modularity}
The shrinker is structured as a collection of local and independent operators, each targeting a specific aspect of the PTA. This modular design supports parsimony through fine-grained elimination of unnecessary components without globally restructuring the model, and allows operators to be combined flexibly.

\subsection{Shrinking Procedure}
The shrinker applies a staged, deterministic shrinking pipeline, where each stage enforces parsimony at a different structural level. Each stage is applied repeatedly until no further shrinking steps are found; the procedure does not backtrack or reorder stages. The overall procedure is as follows:

\begin{enumerate}
    \item \textbf{Transition Removal.} Transitions are removed one at a time, accepting a removal only if the target property is preserved.
    \item \textbf{Reachability Normalization.} After transition removals, all unreachable locations are deterministically eliminated (not property-dependent).
    \item \textbf{Invariant and Guard Simplification.} Guards and invariants are simplified by removing conjuncts, provided the target property remains satisfied.
    \item \textbf{Location Coalescing.} Locations are merged when possible while preserving well-formedness and satisfaction of the target property.
    \item \textbf{Variable Cleanup.} Unused clocks and parameters (not appearing in guards, invariants, or resets) are removed in a final cleanup step (not property-dependent).
\end{enumerate}

\section{Implementation and Evaluation}
\label{sec:evaluation}
We instantiate the framework as a differential testing tool~\cite{DBLP:journals/dtj/McKeeman98} for PTA analysis algorithms: two configurations of the same algorithm---or two algorithms whose outputs should agree on a shared model class---are run on each generated PTA, and disagreement serves as the target property $\varphi$.
Concretely, given two analysis procedures $\Pi_1$ and $\Pi_2$, we define
\[
\varphi(\automaton) \;\triangleq\; \Pi_1(\automaton) = \Pi_2(\automaton).
\]
Whenever $\varphi$ is violated, the shrinking procedure reduces the witness to a local minimum, yielding a small and actionable counterexample.
This strategy is especially effective for testing optimizations and implementation-specific heuristics: if $\Pi_2$ is a trusted reference and $\Pi_1$ an optimized variant, any counterexample to $\varphi$ pinpoints a semantic inconsistency introduced by the optimization.

\subsection{Implementation Details}
The framework is implemented as a meta-tool within the \imitator model checker~\cite{DBLP:conf/cav/Andre21}. The generator is realized as a custom random PTA generator adhering to the design principles of \cref{sec:generator}, while the reducer implements the property-preserving shrinking procedure described in \cref{sec:shrinker}.\footnote{Available at \href{https://github.com/imitator-model-checker/imitator/tree/feature/property-based-testing}{github.com/imitator-model-checker/imitator}} 

\imitator provides multiple PTA analysis procedures; in this section, we focus on three relatively recent additions: a parametric timed game (PTG) algorithm~\cite{dahlsenjensen2024ontheflyalgorithmreachabilityparametric}, currently under active development; an unavoidability (AF) synthesis algorithm~\cite{DBLP:conf/rp/AndreLR15}, an experimental implementation of an established theory; and an NDFS algorithm~\cite{DBLP:conf/iceccs/NguyenPP18} for detecting accepting cycles, a mature and well-tested procedure. All three expose numerous toggleable optimizations and features, making them suitable candidates for differential validation. 

\subsection{Experimental Setup}

All experiments were conducted on a single core of a computer with an Intel Core
i5-1135G7 CPU @ 2.40GHz with 16 GB of RAM running Ubuntu 22.04.5 LTS.
Each experiment ran with $r = 100{,}000$ random iterations per seed, repeated over 5 seeds, with a per-run time limit of 0.1 seconds.
Each experiment compares two configurations of \imitator on the same randomly generated PTA; a disagreement in their outputs triggers the shrinking procedure.
Seeds that did not find a counterexample within $r$ iterations are included in the average count with a value of $r$, giving a conservative estimate.
\Cref{tab:differential} summarizes the results.

If a bug was found and later fixed, the experiment is run on an older version of \imitator and the counterexample is confirmed as fixed on the latest version.
Three experiments---\textbf{PTG-Closed-Clock-Bug}, \textbf{PTG-Explore-Bug}, and \textbf{PTG-Invariant-Bug}---target distinct bugs in the PTG algorithm by comparing the base implementation against a version with a new convex hull abstraction. The bugs themselves were found to be unrelated to the abstraction, but the difference in state space size and exploration order makes each bug manifest in the comparison.
\textbf{PTG-Double-Inclusion-Bug} targets a bug in the double inclusion (two-way subsumption) check of the PTG algorithm, comparing the base implementation against one with the check enabled. This check is still experimental and the bug has not yet been fixed.
\textbf{AF-Bug} compares the PTG and AF synthesis algorithms on the same models, exploiting the fact that their outputs should agree on a subclass of PTA; it finds a bug in the AF algorithm that has not yet been fixed.
The final two experiments, \textbf{NDFS-Check} and \textbf{AF-Double-Inclusion-Check}, serve as negative controls, comparing two configurations of the same algorithm that should produce identical outputs.

\begin{table}[t]
\centering
\small
\adjustbox{max width=\linewidth}{%
\begin{tabular}{l c c c c c}
\toprule
Experiment  & Seeds  & Avg.\ count  & Reduction  & Final size  & Fixed \\
\midrule

PTG-Closed-Clock-Bug
  & 4/5
  & 58624
  & 21\% / 48\%
  & 4 / 4
  & Yes \\

PTG-Explore-Bug
  & 3/5
  & 68081
  & 36\% / 54\%
  & 5 / 6
  & Yes \\

PTG-Double-Inclusion-Bug
  & 5/5
  & 853
  & 20\% / 60\%
  & 3 / 5
  & No \\

PTG-Invariant-Bug
  & 3/5
  & 54374
  & 42\% / 66\%
  & 4 / 4
  & Yes \\

AF-Bug
  & 5/5
  & 14
  & 67\% / 93\%
  & 2 / 1
  & No \\

NDFS-Check
  & 0/5
  & --
  & --
  & --
  & -- \\

AF-Double-Inclusion-Check
  & 0/5
  & --
  & --
  & --
  & -- \\

\bottomrule
\end{tabular}
}
\caption{Results of differential validation experiments on \imitator. Columns:
``Avg.\ count'' = mean PTAs generated before a counterexample;
``Seeds'' = seeds (of 5) finding a witness;
``Reduction'' = percentage reduction in locations/transitions after shrinking;
``Final size'' = shrunk counterexample size in locations/transitions;
``Fixed'' = confirmed fixed in latest \imitator.}
\label{tab:differential}
\end{table}

\subsection{Discussion}
The results confirm that the framework reliably exposes semantic disagreements across a range of algorithm configurations.
The wide variance in average PTA count (14--68081) reflects differences in bug rarity: \textbf{AF-Bug} is triggered by a common structural pattern, whereas \textbf{PTG-Explore-Bug} requires a more specific combination of features.
The shrinking procedure consistently produces substantial reductions, removing up to 93\% of transitions and 67\% of locations.
The two bugs marked as not yet fixed (\textbf{PTG-Double-Inclusion-Bug} and \textbf{AF-Bug}) were confirmed present in the current \imitator release at time of writing, and the generated counterexamples are small and informative enough to be directly actionable for debugging. For instance, the \textbf{AF-Bug} counterexample consists of a PTA with only 2 locations and 1 transition and is most likely an unhandled edge case. 
The two negative-control experiments produced no counterexamples, confirming the absence of spurious disagreements.

\section{Conclusion}
\label{sec:conclusion}
The experiments confirm that analysis algorithms under active development harbour corner-case bugs that standard benchmark suites do not exercise.
The proposed generate-and-shrink framework surfaces such bugs reliably and
reduces any witness to a small, actionable counterexample; the smallest found
consists of just two locations and one transition, yet exposes a fault that
had gone undetected in \imitator until now.

The framework has two main limitations.
First, differential testing requires that the algorithms under comparison
produce deterministic output; non-deterministic implementations may yield
spurious disagreements.
Second, the shrinking procedure finds a locally minimal witness rather than
a globally minimal one: a smaller counterexample may exist.
The framework is most naturally positioned as a development tool --- cheap to run and requiring no oracle --- and extends naturally to other classes of quantitative automata, which we leave to future work.

\appendix

\bibliographystyle{plain}
\bibliography{bibliography.bib}

\begin{thebibliography}{10}

\bibitem{DBLP:conf/stoc/AlurHV93}
Rajeev Alur, Thomas~A. Henzinger, and Moshe~Y. Vardi.
\newblock Parametric real-time reasoning.
\newblock In {\em {STOC}}, pages 592--601. {ACM}, 1993.

\bibitem{DBLP:conf/cav/Andre21}
{\'{E}}tienne Andr{\'{e}}.
\newblock {IMITATOR} 3: Synthesis of timing parameters beyond decidability.
\newblock In {\em {CAV} {(1)}}, volume 12759 of {\em Lecture Notes in Computer
  Science}, pages 552--565. Springer, 2021.

\bibitem{DBLP:conf/rp/AndreLR15}
{\'{E}}tienne Andr{\'{e}}, Didier Lime, and Olivier~H. Roux.
\newblock Integer-complete synthesis for bounded parametric timed automata.
\newblock In {\em {RP}}, volume 9328 of {\em Lecture Notes in Computer
  Science}, pages 7--19. Springer, 2015.

\bibitem{DBLP:conf/tap/AndreMP21}
{\'{E}}tienne Andr{\'{e}}, Dylan Marinho, and Jaco van~de Pol.
\newblock A benchmarks library for extended parametric timed automata.
\newblock In {\em TAP@STAF}, volume 12740 of {\em Lecture Notes in Computer
  Science}, pages 39--50. Springer, 2021.

\bibitem{10.1145/1670412.1670413}
Robert Brummayer and Armin Biere.
\newblock Fuzzing and delta-debugging smt solvers.
\newblock In {\em Proceedings of the 7th International Workshop on
  Satisfiability Modulo Theories}, SMT '09, page 1--5, New York, NY, USA,
  2009. Association for Computing Machinery.

\bibitem{DBLP:conf/icfp/ClaessenH00}
Koen Claessen and John Hughes.
\newblock Quickcheck: a lightweight tool for random testing of haskell
  programs.
\newblock In {\em {ICFP}}, pages 268--279. {ACM}, 2000.

\bibitem{dahlsenjensen2024ontheflyalgorithmreachabilityparametric}
Mikael~Bisgaard Dahlsen-Jensen, Baptiste Fievet, Laure Petrucci, and Jaco
  van~de Pol.
\newblock On-the-fly algorithm for reachability in parametric timed games
  (extended version), 2024.

\bibitem{DBLP:conf/tase/ManiniRP25}
Andrea Manini, Matteo~G. Rossi, and Pierluigi {San Pietro}.
\newblock Random testing of model checkers for timed automata with automated
  oracle generation.
\newblock In {\em {TASE}}, volume 15841 of {\em Lecture Notes in Computer
  Science}, pages 343--360. Springer, 2025.

\bibitem{DBLP:journals/dtj/McKeeman98}
William~M. McKeeman.
\newblock Differential testing for software.
\newblock {\em Digit. Tech. J.}, 10(1):100--107, 1998.

\bibitem{DBLP:conf/iceccs/NguyenPP18}
Hoang~Gia Nguyen, Laure Petrucci, and Jaco van~de Pol.
\newblock Layered and collecting {NDFS} with subsumption for parametric timed
  automata.
\newblock In {\em {ICECCS}}, pages 1--9. {IEEE} Computer Society, 2018.

\bibitem{DBLP:conf/pldi/RegehrCCEEY12}
John Regehr, Yang Chen, Pascal Cuoq, Eric Eide, Chucky Ellison, and Xuejun
  Yang.
\newblock Test-case reduction for {C} compiler bugs.
\newblock In {\em {PLDI}}, pages 335--346. {ACM}, 2012.

\bibitem{DBLP:conf/lpar/TabakovV05}
Deian Tabakov and Moshe~Y. Vardi.
\newblock Experimental evaluation of classical automata constructions.
\newblock In {\em {LPAR}}, volume 3835 of {\em Lecture Notes in Computer
  Science}, pages 396--411. Springer, 2005.

\bibitem{DBLP:conf/pldi/YangCER11}
Xuejun Yang, Yang Chen, Eric Eide, and John Regehr.
\newblock Finding and understanding bugs in {C} compilers.
\newblock In {\em {PLDI}}, pages 283--294. {ACM}, 2011.

\bibitem{DBLP:conf/esec/Zeller99}
Andreas Zeller.
\newblock Yesterday, my program worked. today, it does not. why?
\newblock In {\em {ESEC} / {SIGSOFT} {FSE}}, volume 1687 of {\em Lecture Notes
  in Computer Science}, pages 253--267. Springer, 1999.

\end{thebibliography}

\newpage
\appendix

\end{document}